\documentstyle[aps,preprint]{revtex}  
\begin{document}                
\preprint{SNUTP 96-032}
\title{Remarks on Renormalization of Black Hole Entropy}
\author{Sang Pyo Kim$^a$ \footnote{Electronic
address: sangkim@knusun1.kunsan.ac.kr},
Sung Ku Kim$^b$ \footnote{Electronic address:
skkim@theory.ewha.ac.kr},
Kwang-Sup Soh$^c$ \footnote{Electronic address:
kssoh@phyb.snu.ac.kr},
and Jae Hyung Yee$^d$ \footnote{Electronic address:
jhyee@phya.yonsei.ac.kr}}

\address{$^a$ Department of Physics,
Kunsan National University,
Kunsan 573-701, Korea \\
$^b$ Department of Physics,
Ewha Womans University,
Seoul 120-750, Korea\\
$^c$ Department of Physics Education,
Seoul National University,
Seoul 151-742, Korea\\
$^d$ Department of Physics
and Institute for Mathematical
Sciences,
Yonsei University,\\
Seoul 120-749, Korea}

\maketitle
\begin{abstract}                
We elaborate  the renormalization process of
entropy of a nonextremal and an extremal
Reissner-Nordstr\"{o}m black hole by
using the Pauli-Villars regularization method, in which
the regulator fields obey either the Bose-Einstein or Fermi-Dirac
distribution depending on their spin-statistics.
The black hole entropy involves only
two renormalization constants.
We also discuss the entropy and temperature of the extremal
black hole.
\end{abstract}

\section{Introduction}

Bekenstein \cite{bekenstein}
and Hawking \cite{hawking} proposed
an intrinsic entropy and temperature of black holes.
The black hole entropy and temperature are purely
geometric in that the former is proportional to
the area of event horizon of each black hole
and the latter is determined by the surface
gravity at the event horizon. Entropy is, however,
a thermodynamic quantity that is usually defined
in terms of the classical or
quantum mechanical number of states available to a system
under consideration.
The geometric nature of black hole entropy
has thus raised the fundamental questions on the
origin of black hole
entropy over the last twenty years.

There have been various attempts to explain the black
hole entropy thermodynamically.
't Hooft calculated the entropy of a scalar field
in a black hole background and found that
with a particular brick wall just outside
the event horizon the thermodynamic
entropy of the scalar field gives the correct
geometric black hole entropy \cite{thooft}.
Quite recently Demers, Lafrance, and Myers (DLM)
introduced a Pauli-Villars method
to regularize the ultraviolet divergences of a
Reissner-Nordstr\"{o}m (RN) black hole entropy \cite{demers}.
In the Pauli-Villars
covariant regularization method,
one introduces bosonic and fermionic
regulator fields to regularize the divergences.
The strong point of the method of Ref. \cite{demers} is that
it does not necessitate a brick wall
to avoid possible ultraviolet divergences
and yields a one-loop
renormalized entropy.
They obtained the thermodynamic entropy of both
the nonextremal and extremal black holes which involves
only two renormalization constants. It was further
shown that these two constants also appeared
in the one-loop effective action of the scalar field
coupled to gravity, and can be absorbed into
the renormalization of the gravitational constant
$G$ and the coupling constants of higher order
curvatures in the effective action.
The renormalization of black hole entropy in terms of
the gravitational constant and the coupling constants
was first discussed in \cite{susskind}
and also discussed in
\cite{soludkhin,alwis,cognola,barbon,larsen,belgiorno}.

As usual, DLM
used somewhat {\it ad hoc} arbitrary spin-statistics,
the Bose-Einstein distribution,
even for the fermionic regulator fields.
In this paper we show that with the right
spin-statistics for the regulator fields the Pauli-Villars
covariant regularization method
can still work to yield a one-loop renormalized
black hole entropy.
We explicitly calculate the thermodynamic entropy
of both the nonextremal and extremal RN black holes.
The thermodynamic entropy computed with this method
is shown to
be of the same form as those in \cite{demers},
but the two renormalization constants involved differ from
those in \cite{demers}.

Throughout this paper we adopt the
units, $c = G = \hbar = k = 1$.
The spacetime signature is $(-, +, +, +)$.

\section{The Entropy of Reissner-Nordstr\"{o}m Black Hole}

The thermodynamic entropy of
a massive scalar field in an RN black hole
background
will be studied in this section. We shall
follow mostly the method used by 't Hooft
\cite{thooft}.
The RN black hole has the metric
\begin{equation}
 ds^2 = -
 \left(1 -
 \frac{r_-}{r} \right)
 \left(1 - \frac{r_+}{r} \right)dt^2
 +
 \frac{dr^2}{\left(1 -
 \frac{r_-}{r} \right)
 \left(1 - \frac{r_+}{r} \right)}
+ r^2 d\Omega^2,
\end{equation}
where $r_{\pm} = M \pm \sqrt{M^2 - Q^2}$,
and $r_+ > r_-$ for the nonextremal RN
black hole
and $ r_+ = r_-$ for the extremal
RN black hole.
The limiting case of $r_- = 0$ and $r_+ = 2M$
corresponds to the Schwarzschild black hole.

We consider a massive scalar field
minimally coupled to the RN black hole background.
The scalar field satisfies the (quantum)
Klein-Gordon equation
\begin{eqnarray}
\Biggl[
- \frac{1}{\bigl(1 - \frac{r_-}{r} \bigr)
\bigl(1 - \frac{r_+}{r} \bigr)}
\frac{\partial^2}{\partial t^2}
&+&
\frac{1}{r^2}
\frac{\partial}{\partial r}
\Bigl( \bigl(r - r_- \bigr)
\bigl(r - r_+ \bigr)
\frac{\partial}{\partial r} \Bigr)
\nonumber\\
&+& \frac{1}{r^2} \Bigl( \frac{1}{\sin\theta}
\frac{\partial}{\partial \theta}
\bigl(\sin \theta \frac{\partial}{\partial \theta}
\bigr) + \frac{\partial^2}{\partial \phi^2}
\Bigr)
 - m^2 \Biggr]
\phi (x) = 0.
\label{kg eq}
\end{eqnarray}
Since the RN black hole has
a timelike Killing vector field
and the spherical symmetry,
we expand the wave function by spherical
harmonics,
$\phi(x) = e^{-i Et} Y_{lm}(\theta, \phi)
f(r)$.
The remaining radial wave function
of the Klein-Gordon equation
satisfies the one-dimensional Schr\"{o}dinger-like equation
\begin{equation}
\Biggl[
\frac{1}{r^2}
\frac{d}{d r}
\Bigl( \bigl(r - r_- \bigr)
\bigl(r - r_+ \bigr)
\frac{d}{d r} \Bigr)
+ \frac{r^2 E^2}{\bigl(r - r_- \bigr)
\bigl(r - r_+ \bigr)}
- \frac{l (l+1)}{r^2} - m^2
\Biggr] f(r) = 0.
\end{equation}
The radial momentum,
$p_r = \frac{d S}{d r}$,
can be found from the WKB
wave function, $f(r) = e^{iS(r)}$,
\begin{equation}
p_r^2 =
\frac{1}{\left(1 - \frac{r_-}{r} \right)^2
\left(1 - \frac{r_+}{r} \right)^2 }
\left[E^2 - \left(1 - \frac{r_-}{r} \right)
\left(1 - \frac{r_+}{r} \right)
\left( \frac{l(l+1)}{r^2} + m^2
\right) \right].
\label{ra mo}
\end{equation}
The scalar field suffers an infinite gravitational
redshift when it passes from the event horizon to
infinity.

We find the Helmholtz free energy from
the number of states available and the statistical
distribution.
For a given $E$ we count the action
in the unit of
Planck constant\footnote{Here we adopted the convention
of action $ \int dr p_r $
and the number of states
$\frac{1}{2\pi \hbar} \int dr p_r$
as in Ref. \cite{goldstein}.}
in order to get the number of states
summed over
the angular momentum states
\begin{eqnarray}
g(E) = &&
\frac{1}{2 \pi} \int_{0}^{\infty} dl (2l + 1)
\int dr
\frac{1}{\left(1 - \frac{r_-}{r} \right)
\left(1 - \frac{r_+}{r} \right) }
\nonumber\\
&& ~~~~~\times
\left[E^2 - \left(1 - \frac{r_-}{r} \right)
\left(1 - \frac{r_+}{r} \right)
\left( \frac{l(l+1)}{r^2} + m^2
\right) \right]^{1/2}.
\label{nu st}
\end{eqnarray}

\subsection{Ultraviolet Divergence}

The radial momentum goes to infinity
at the event horizon $r = r_+$
and it causes an ultraviolet divergence.
To avoid the ultraviolet divergences
't Hooft introduced
a brick wall of thickness $h$
just outside the event horizon
to regularize the divergences.
The number of states after
performing the angular momentum integration
amounts to
\begin{equation}
g(E) =
\frac{1}{3 \pi}
\int_{r_+ + h} dr
\frac{r^2}{\left(1 - \frac{r_-}{r} \right)^2
\left(1 - \frac{r_+}{r} \right)^2 }
\left[E^2 - \left(1 - \frac{r_-}{r} \right)
\left(1 - \frac{r_+}{r} \right) m^2 \right]^{3/2},
\label{nu st1}
\end{equation}
and the Helmholtz free energy is
\begin{equation}
F =  \frac{1}{\beta} \sum_{N}
\ln \left(1 - e^{- \beta E_N} \right).
\end{equation}
For a continuous $E$,
we may take a continuum limit to compute
the Helmholtz free energy
\begin{equation}
F = - \int_{0}^{\infty}
dE  g(E) \frac{1}{ e^{ \beta E} - 1}.
\label{he en}
\end{equation}
Substituting (\ref{nu st}) into (\ref{he en})
and changing the variable to $x = 1 - \frac{r_+}{r}$,
we finally obtain
\begin{equation}
F_{RN} = - \frac{r_+^3}{3 \pi}
\int_0^{\infty} dE \frac{1}{e^{\beta E} - 1}
\int_{\epsilon}
dx \frac{1}{x^2 (1 - x)^4 (1 - u + ux)^2}
\left[E^2 - m^2 x(1 - u + ux) \right]^{3/2}
\label{in fr}
\end{equation}
where
$u = \frac{r_-}{r_+}$, $\epsilon = \frac{h}{r_+ + h}$,
and
$u =1$ corresponds to the  exremal RN black hole.

We focus on the possible ultraviolet
divergences coming from the event horizon
and expand the rational
function part of integral (\ref{in fr})
around $ x = 0$. The
Helmholtz free energy
for the nonextremal black hole
is then given by
\begin{equation}
F_{n.ext} = - \frac{r_+^3}{3 \pi} \sum_{n = 0}^{\infty}
C_n
\int_{0}^{\infty} dE \frac{1}{e^{\beta E} - 1}
 \int_{\epsilon} dx   x^{-2 + n}
\left[E^2 - m^2 x(1-u +ux) \right]^{3/2}
\label{no fr}
\end{equation}
where
\begin{equation}
C_0 = \frac{1}{(1 - u)^2}, ~
C_1 =  \frac{2(2 -3u)}{(1 - u)^3},
\end{equation}
and for the extremal black hole
by
\begin{equation}
F_{ext} = - \frac{r_+^3}{3 \pi} \sum_{n = 0}^{\infty}
D_n
\int_0^{\infty} dE \frac{1}{e^{\beta E} - 1}
\int_{\epsilon} dx
x^{-4 + n} \left[E^2 - m^2 x^2 \right]^{3/2},
\label{ex fr}
\end{equation}
where
\begin{equation}
D_0 = 1, ~~D_n = \frac{4 \cdot 5 \cdots (3 + n)}{n!} ,
(n \geq 1).
\end{equation}

We show explicitly how the ultraviolet divergent
terms in the nonextremal black hole
emerge by performing integration of
the first two terms in (\ref{no fr}).
By keeping only the divergent terms \cite{grad},
we have
\begin{eqnarray}
F_{n.ext} \sim - \frac{r_+^3}{3 \pi}
\Biggl[ \frac{\pi^4}{15 (1-u)^2 \beta^4}
\frac{1}{\epsilon}
+ \frac{\pi^2 m^2}{8 (1 - u) \beta^2}
\ln (\epsilon^2)
- \frac{\pi^4 (2 - 3u)}{15 (1 - u)^3
\beta^4} \ln (\epsilon^2)
\Biggr],
\label{on fr1}
\end{eqnarray}
where we used the integral formula, for $n = 0,1$
of (\ref{no fr}),
\begin{eqnarray}
&&\int dx \frac{1}{x \sqrt{E^2 - m^2 x(1 - u + ux)}}
= \frac{1}{E} \ln \left(
\frac{\sqrt{ E + a_+x} - \sqrt{E + a_-x}}{\sqrt{E + a_+ x}
+ \sqrt{E + a_- x}}\right),
\nonumber\\
&&a_{\pm} = \frac{- (1 -u)m^2 \pm \sqrt{(1 - u)^2 m^4
+ 4 u m^2 E^2}}{2E}.
\label{in1}
\end{eqnarray}
We note that $F_{n.ext}$   contains
only a linearly divergent term $\frac{E^3}{\epsilon}$
and logarithmically divergent terms $E^3 \ln (\epsilon^2)$
and $E m^2 \ln ( \epsilon^2)$ as $\epsilon \rightarrow 0$.
The Helmholtz free energy of the scalar field thus diverges
as the brick wall thickness $h$ (here $\epsilon$) collapses to
zero at the event horizon.

The ultraviolet divergent terms
in the extremal black hole
similarly come from the first
four terms in (\ref{ex fr}).
Keeping only the divergent terms
we get
\begin{equation}
F_{ext} \sim - \frac{r_+^3}{3 \pi}
\Biggl[ \frac{\pi^4}{45 \beta^4}
\frac{1}{\epsilon^3}
+ \frac{2 \pi^4}{15 \beta^4}
\frac{1}{\epsilon^2}
+ \frac{2 \pi^4}{3 \beta^4}
\frac{1}{\epsilon}
- \frac{\pi^2 m^2}{6 \beta^2}
\frac{1}{\epsilon}
- \frac{2 \pi^4}{3 \beta^4}
\ln (\epsilon^2)
+ \frac{\pi^2 m^2}{2 \beta^2}
\ln (\epsilon^2) \Biggr],
\end{equation}
where the use has been made of the following
integral formula \cite{grad} , for $n = 1, 3$ of
(\ref{ex fr})
\begin{equation}
\int dx \frac{1}{x \sqrt{E^2 - m^2 x^2}}
= \frac{1}{2E} \ln \left(\frac{E - \sqrt{E^2 - m^2 x^2}}{
E + \sqrt{E^2 - m^2 x^2}} \right).
\label{in2}
\end{equation}
The divergent terms are $\frac{E^3}{\epsilon^3}$,
$ \frac{E^3}{\epsilon^2}$, $\frac{E^3}{\epsilon}$,
$\frac{E m^2}{\epsilon}$, $ E^3 \ln (\epsilon^2)$,
and $ E m^2 \ln (\epsilon^2)$.

The main purpose of this paper is to see
whether one may remove
these divergent terms and may renormalize the free energy
by introducing bosonic and fermionic
regulator fields that obey
the proper Bose-Einstein or Fermi-Dirac
statistics. This point will be discussed
in Sec. III.

\subsection{Brick Wall Model}

't Hooft introduced a brick wall just outside
the event horizon to cut off the ultraviolet
divergence and suggested that the thermodynamic
entropy of scalar fields might give the correct
geometric entropy of black hole if a particular brick
wall thickness is chosen. We reconsider his
argument using the entropy
of the nonextremal RN black hole
including the logarithmically divergent terms.

First, the Schwarzchild black hole case is recovered
from the limiting case of $r_- = 0$ and $r_+ = 2M$.
Repeating the calculation,
focusing only on the dominant divergent terms
of the Helmholtz free energy,
and using
$S = \beta^2 \frac{\partial F}{\partial \beta}$,
we find the thermodynamic entropy
\begin{equation}
S_{Sch} \sim  \frac{r_+^3}{3 \pi}
\Biggl[ \frac{4\pi^4}{15 \beta^3}
\frac{1}{\epsilon}
+ \frac{\pi^2 m^2}{2 \beta}
\ln (\epsilon)
- \frac{16 \pi^4}{15 \beta^3} \ln (\epsilon)
\Biggr].
\label{sc en}
\end{equation}
't Hooft ascribed the black hole entropy
only to the linearly divergent term
and prescribed
\begin{equation}
\epsilon = \frac{1}{180 A},
\end{equation}
where $A = 16\pi M^2$ is the area of the event
horizon of Schwarzchild black hole.
He thus obtained $S \sim \frac{A}{4}$.
However, assigning $\beta = 8\pi M$
one gets the on-shell black hole entropy
\begin{equation}
S_{Sch} \sim \frac{1}{720} \frac{1}{\epsilon}
+ \frac{m^2 A}{96 \pi} \ln(\epsilon)
 - \frac{1}{180} \ln(\epsilon).
\end{equation}
It is to be noted that the
second term of (\ref{sc en})
gives rise to an entropy proportional to
the area of the event horizon, so this term
should not be neglected even for small $\epsilon$.
We should instead prescribe
\begin{equation}
\epsilon \left(1 - \frac{m^2}{24 \pi}
\ln(\epsilon) \right) =
\frac{1 - 4 \epsilon \ln(\epsilon)}{180 A}
\end{equation}
to get the black hole entropy
\begin{equation}
S_{Sch} \sim \frac{A}{4}.
\end{equation}

Similarly, we find the thermodynamic entropy of
the nonextremal RN black hole
\begin{equation}
S_{n.ext} \sim  \frac{r_+^3}{3 \pi}
\Biggl[ \frac{4 \pi^4}{15 (1-u)^2 \beta^3}
\frac{1}{\epsilon}
+ \frac{\pi^2 m^2}{2(1 - u) \beta}
\ln (\epsilon)
- \frac{8 \pi^4 (2 - 3u)}{15 (1 - u)^3
\beta^3} \ln (\epsilon)
\Biggr].
\label{nex ent}
\end{equation}
The on-shell entropy, $\beta = \frac{4\pi r_+}{1 - u}$, is
\begin{equation}
S_{n.ext} \sim \frac{1 - u}{720} \frac{1}{\epsilon}
+ \frac{m^2 A}{96 \pi} \ln(\epsilon)
 - \frac{2- 3 u}{360} \ln(\epsilon).
\end{equation}
By prescribing the brick wall condition
\begin{equation}
\epsilon \left(1 - \frac{m^2}{24 \pi}
\ln(\epsilon) \right) =
\frac{(1 - u)(1 - 4 \epsilon \ln(\epsilon))}{180 A},
\label{br wa}
\end{equation}
we obtain the black hole entropy
\begin{equation}
S_{n.ext} \sim \frac{A}{4}.
\end{equation}

Thus we have seen that
the same entropy formula is obtained by
including the logarithmically  divergent term
that 't Hooft neglected.
The contribution of logarithmically
divergent term to the 2D black hole
entropy was discussed in \cite{soludkhin}.

\section{Pauli-Villars Regularization}

It is our observation that by using the correct
Bose-Einstein
and Fermi-Dirac statistics for the regulator fields
we
can remove the same divergent terms for
both the nonextremal and
extremal RN black hole,
because they contribute to the Helmholtz free energy
with opposite signs, and
the Bose-Einstein distribution
\begin{equation}
\int_0^{\infty} dE \frac{E^{\nu -1}}{e^{\beta E} - 1}
= \frac{\Gamma (\nu) \zeta (\nu)}{\beta^{\nu}}
\end{equation}
and the Fermi-Dirac distribution
\begin{equation}
\int_0^{\infty} dE \frac{E^{\nu -1}}{e^{\beta E} + 1}
= \left(1 - 2^{1 - \nu} \right)
\frac{\Gamma (\nu) \zeta (\nu)}{\beta^{\nu}}
\end{equation}
differ by only a constant factor.
This suggests
that in order to remove the ultraviolet
divergences we may introduce
bosonic and fermionic
regulator fields
that satisfy the Klein-Gordon equation (\ref{kg eq})
with arbitrarily large masses $m_i$.
The $m_i$ will be determined later
from the mass conditions that make all the ultraviolet
divergent terms cancel each other.

Repeating the same calculation as for
the bosonic massive scalar field,
we get the number of states for each regulator field
\begin{equation}
g_i(E) =
\frac{1}{3 \pi}
\int_{r_+ + h} dr
\frac{r^2}{\left(1 - \frac{r_-}{r} \right)^2
\left(1 - \frac{r_+}{r} \right)^2 }
\left[E^2 - \left(1 - \frac{r_-}{r} \right)
\left(1 - \frac{r_+}{r} \right) m^2_i \right]^{3/2},
\end{equation}
where $i$ labels the species of regulator fields.
Similarly, we obtain the Helmholtz free energy
contribution of the $i$th regulator field
\begin{equation}
F_i =  \frac{1}{\beta} \sum_{N}
\ln \left(1 \mp e^{- \beta E_N} \right),
\end{equation}
and for continuous $E$
\begin{equation}
F_i = \mp \frac{r_+^3}{3 \pi}
\int_0^{\infty} dE \frac{1}{e^{\beta E} \mp 1}
\int_{\epsilon}
dx \frac{1}{x^2 (1 - x)^4 (1 - u + ux)^2}
\left[E^2 - x(1 - u + ux)m^2_i \right]^{3/2},
\label{in fr1}
\end{equation}
where the upper (-) sign is used for bosonic fields and
the lower (+) sign for fermionic fields,
and we used the same variables and parameters as in
Sec. II. As before, $u = 1$ corresponds
to the extremal black hole case.

\subsection{Nonextremal RN Black Hole}

The ultraviolet divergent terms
of each regulator fields in the nonextremal RN black hole
come from the singular behavior of
the Helmholtz free energy (\ref{in fr1})
near $x = 0$.
We compute the Helmholtz free energy around $ x = 0$:
\begin{equation}
F_i = \mp \frac{r_+^3}{3 \pi} \sum_{n = 0}^{\infty}
C_n
\int_{0}^{\infty} dE \frac{1}{e^{\beta E} \mp 1}
 \int_{\epsilon} dx   x^{-2 + n}
\left[E^2 - m_i^2 x(1-u +ux) \right]^{3/2}
\end{equation}
where
\begin{equation}
C_0 = \frac{1}{(1 - u)^2}, ~
C_1 =  \frac{2(2 -3u)}{(1 - u)^3}.
\end{equation}
Regardless of spin-statistics of regulator fields,
the free energy contains
a linearly divergent term $\frac{E^3}{\epsilon}$
and logarithmically divergent terms $E^3 \ln (\epsilon^2)$
and $E m^2_i \ln (\epsilon^2)$
as $\epsilon \rightarrow 0$ as shown in Sec. II.
We may remove these divergent terms
by introducing the correct bosonic and fermionic
regulator fields that contribute to the free energy
(\ref{in fr1}) with opposite signs,
and thereby regularize the free energy and entropy.

The linearly divergent contribution to the free energy is
\begin{equation}
- \frac{1}{\epsilon} \left(N_B  - \frac{7}{8}
N_F \right) \frac{\pi^3 r_+^3}{45 (1-u)^2 \beta^4},
\end{equation}
where $N_B$ and $N_F$ are the number of the bosonic
and fermionic fields, respectively.
We can remove this term by introducing $7k$ ($k$ a positive
integer) bosonic fields (including
the original scalar field)
and $8k$ fermionic regulator fields. In this paper
we shall take the minimum number of regulator fields,
that is, 7 bosonic and 8 fermionic fields.

Next, we consider the logarithmically divergent contributions
\begin{eqnarray}
\left(N_B  - \frac{7}{8}
N_F \right) \frac{\pi^3 (2 - 3u) r_+^3}{45 (1-u)^3 \beta^4}
\ln (\epsilon^2),
\nonumber\\
- \left(\sum_{B}^{7} m^2_i
- \frac{1}{2} \sum_{F}^{8} m^2_i \right)
\frac{\pi r_+^3}{24 (1 - u) \beta^2}
\ln (\epsilon^2).
\end{eqnarray}
If we further impose a condition
on the masses of the regulator fields
\begin{equation}
\sum_{B}^{7} m^2_i - \frac{1}{2} \sum_{F}^{8} m^2_i = 0,
\label{ma co}
\end{equation}
then both the linearly divergent term and
logarithmically divergent terms are made vanish.
So we may remove the brick wall, $\epsilon = 0$,
under the condition (\ref{ma co}),
and instead regularize the free energy by the
large masses, $m_i$, of regulator fields.

We find the Helmholtz free energy
in the limit $\epsilon = 0$. For this purpose
we expand the right hand side of the integral (\ref{in1}),
use the mass condition (\ref{ma co}) to remove
$\ln (\epsilon^2)$,
and obtain the only nonvanishing integral
\begin{equation}
\int_0 dx \frac{1}{x \sqrt{E^2 - x(1 - u + ux)m^2}}
= \frac{1}{2E} \ln \left(
\frac{(1 - u)^2 m_i^4 + 4 u m_i^2 E^2}{16 E^4} \right).
\label{in3}
\end{equation}
Then the Helmholtz free energy is given by
\begin{eqnarray}
F_{n.ext} = - \frac{r_+^3}{3 \pi}
\Biggl[&&
\frac{3}{4(1 -u)} \sum_B^7 m_i^2
\int_0^{\infty} dE \frac{E \ln\Bigl(
\frac{(1-u)^2 m_i^4 + 4 u m_i^2 E^2}{16E^4}
\Bigr)}{e^{\beta E} -1}
\nonumber\\
&&-
\frac{3}{4(1 -u)} \sum_F^8 m_i^2
\int_0^{\infty} dE \frac{E \ln\Bigl(
\frac{(1-u)^2 m_i^4 + 4 u m_i^2 E^2}{16E^4}
\Bigr)}{e^{\beta E} + 1}
\nonumber\\
&&-
\frac{2 - 3u}{(1 -u)^3} \sum_B^7
\int_0^{\infty} dE \frac{E^3 \ln\Bigl(
\frac{(1-u)^2 m_i^4 + 4 u m_i^2 E^2}{16E^4}
\Bigr)}{e^{\beta E} -1}
\nonumber\\
&&+
\frac{2 - 3u}{(1 -u)^3} \sum_F^8
\int_0^{\infty} dE \frac{E^3 \ln\Bigl(
\frac{(1-u)^2 m_i^4 + 4 u m_i^2 E^2}{16E^4}
\Bigr)}{e^{\beta E} + 1}
\Biggr].
\label{fr en}
\end{eqnarray}

For large masses of regulator fields, we expand
the logarithmic terms in (\ref{fr en}) in the inverse power of
masses. The leading terms are
$\frac{1}{E} \left(\ln (m_i^2 - 2 \ln (E) \right)$.
Keeping only the surviving divergent contributions to the
Helmholtz free energy as the regulator masses go to
infinity
we obtain
\begin{equation}
F_{n.ext} =  -
\frac{\pi r_+^3 {\cal B}}{12 (1 - u) \beta^2}
- \frac{2\pi^3 r_+^3 (2-3u) {\cal A}}{45 (1 -u)^3 \beta^4},
\label{fr en1}
\end{equation}
where
\begin{eqnarray}
{\cal A} &=&  - \sum_B^7 \ln (m_i^2) + \frac{7}{8}
\sum_F^8 \ln (m_i^2),
\\
{\cal B} &=& \sum_B^7 m_i^2 \ln (m_i^2)
- \frac{1}{2} \sum_F^8 m_i^2 \ln (m_i^2)
\nonumber\\
&&~- \frac{12}{\pi^2}
\sum_B^7 m_i^2
\int_0^{\infty} dt \frac{t\ln(t)}{e^{t} - 1}
+ \frac{12}{\pi^2}  \sum_F^8 m_i^2
\int_0^{\infty} dt  \frac{t\ln(t)}{e^{t} + 1}.
\label{re co}
\end{eqnarray}
From the inequality $t > \ln(t) \geq - \frac{1}{e \alpha t^{\alpha}}$
for $ 1 > \alpha > 0$,
we may find the bound for the integrals of (\ref{re co}):
\begin{eqnarray}
&&\Gamma(3) \zeta(3) > \int_0^{\infty} dt
\frac{t \ln (t)}{e^t - 1}
\geq - \frac{\Gamma(2 - \alpha)
\zeta(2 - \alpha)}{e \alpha},
\nonumber\\
&&\frac{3}{4} \Gamma(3) \zeta(3) > \int_0^{\infty} dt
\frac{t \ln (t)}{e^t + 1}
\geq - \left(1 - \frac{1}{2^{1 - \alpha}} \right)
\frac{\Gamma(2 - \alpha)
\zeta(2 - \alpha)}{e \alpha}.
\end{eqnarray}

The black hole entropy is then
\begin{equation}
S_{n.ext} =
\frac{\pi r_+^3 {\cal B}}{6 (1 - u) \beta}
+ \frac{8 \pi^3 r_+^3 (2-3u) {\cal A}}{45 (1 -u)^3 \beta^3}.
\label{no en}
\end{equation}
Note that if we had used the same
spin-statistics even for
the bosonic and fermionic regulator fields,
the constants ${\cal A}$ and ${\cal B}$ would become
\begin{equation}
{\cal A} = - \sum_B^3 \ln(m_i^2)
+ \sum_F^3 \ln(m_i^2),~~
{\cal B} =  \sum_B^3 m_i^2 \ln(m_i^2)
- \sum_F^3 m_i^2 \ln(m_i^2)
\end{equation}
for three bosonic and fermionic fields, repectively,
which are
the same as those in \cite{demers}.

The entropy obtained so far is off-shell and we substitute
the Hawking temperature,
$\beta = \frac{4 \pi r_+}{1 -u}$,
of the nonextremal RN black hole
to get the on-shell entropy
\begin{equation}
S_{n.ext} =
\frac{{\cal B}}{24 \pi} \frac{A}{4}
+ \frac{(2-3u) {\cal A}}{180},
\label{no en1}
\end{equation}
where $A = 4 \pi r_+^2$ is the surface area of the event horizon.
It should be remarked that we obtained the
exactly same form
of entropy as in \cite{demers}, but  with different
renormalization constants ${\cal A}$ and ${\cal B}$.
${\cal A}$ and ${\cal B}$  might be related with the
renormalization of $G$ and the coefficients of the effective
action \cite{susskind}.

\subsection{Extremal RN Black Hole}

Now we turn to the exremal RN black hole,
the case with $ u = 1$.
The Helmholtz free energy contribution
of the $i$th regulator field is
\begin{equation}
F_i = \mp \frac{r_+^3}{3 \pi \hbar}
\int_0^{\infty} dE \frac{1}{e^{\beta E} \mp 1}
\int_{\epsilon}
dx \frac{1}{x^4 (1 - x)^4}
\left[E^2 - m^2_i x^2 \right]^{3/2}.
\end{equation}
We expand $ \frac{1}{(1 - x)^4}$ around $x = 0$, and
perform the integration
\begin{equation}
F_i = \mp \frac{r_+^3}{3 \pi \hbar} \sum_{n = 0}^{\infty}
D_n
\int_0^{\infty} dE \frac{1}{e^{\beta E} \mp 1}
\int_{\epsilon} dx
x^{-4 + n} \left[E^2 - m^2_i x^2 \right]^{3/2}.
\end{equation}
The divergent terms are $\frac{E^3}{\epsilon^2}$,
$\frac{E m^2_i}{\epsilon}$, $ E^3 \ln (\epsilon^2)$,
and $ E m^2_i \ln (\epsilon^2)$.
Taking the correct spin-statistics of the bosonic and
fermionic regulator fields into account,
the ultraviolet divergences can be removed
with the same condition on the number of regulator fields
and the mass condition (\ref{ma co}) as in the
nonextremal RN black hole. We remove again the brick wall outside
the event horizon, and let $\epsilon = 0$.
Then the nonvanishing contribution to the free 
energy from the lower
limit of integration is
\begin{equation}
F_{ext} =  -
\frac{\pi r_+^3 {\cal B}}{6 \beta^2}
- \frac{2 \pi^3 r_+^3 {\cal A}}{9 \beta^4},
\label{fr en2}
\end{equation}
and the entropy is found to be
\begin{equation}
S_{ext} = \frac{\pi r_+^3 {\cal B}}{3 \beta}
+ \frac{8 \pi^3 r_+^3 {\cal A}}{9 \beta^3}.
\label{ex en}
\end{equation}
The black hole entropy (\ref{ex en}) has also
the same form
as in \cite{demers}, but
differs only  by the renormalization constants
${\cal A}$ and ${\cal B}$.

\section{Conclusion and Discussion}

In this paper, we obtained the thermodynamic
entropy of the RN black holes
using the Pauli-Villars regularization method.
The primary difference of our method from that
of DLM \cite{demers}
is that we used the Bose-Einstein statistics for
the bosonic regulator fields
and the Fermi-Dirac statistics for the
fermionic regulator fields, whereas they
used the same Bose-Einstein statistics for
both the bosonic and fermionic regulator fields.
We confirm that the thermodynamic
entropy of either the nonextremal or
extremal RN black hole involves only two constants
${\cal A}$ and ${\cal B}$ as
in \cite{demers}. But the regulator
masses satisfy the mass condition
(\ref{ma co}) which is different from that in
\cite{demers} due to
the different spin-statistics of regulator fields.
The renormalization constants of
the thermodynamic entropy might be related with
the renormalization of the gravitational
constant $G$ and coupling constants of higher
order curvatures of the effective action.

The thermodynamic entropy of the extremal black hole
has the same form as that of the nonextremal black hole.
There still remains the problem of defining the temperature
and the entropy. From our result we can infer
several possibilities. First, as argued in many literatures
the entropy in (\ref{ex en}) goes to zero as
the temperature approaches to zero. This fact is
consistent with the argument in many
literatures that both temperature
and entropy of the extremal black hole must vanish.
The second possibility is that the temperature is still
inversely proportional to $r_+$ as for the Schwarzschild
or nonextremal RN black holes and the area rule of black
hole entropy remains valid. The third possibility is
that the temperature is arbitrary
\cite{hawking1} but it is related to the
entropy as (\ref{ex en}).

In \cite{demers} it was shown that the renormalization
constants appearing in the entropy
coincide with those from the coupling
constants of the higher order curvatures in the
one-loop effective action.
It would be interesting to find the way to relate
these renormalization constants with the correct
spin-statistics for the regulator fields.

\section*{acknowledgments}
This work was supported in part by the Korea Science and
Engineering Foundation under Grant No. 951-0207-56-2,
95-0701-04-01-3,
and in part by the Basic Science Research Institute Program,
Ministry of Education
under Project No. BSRI-96-2418, BSRI-96-2425, BSRI-96-2427,
and by the Center for Theoretical
Physics, Seoul National University.

\end{document}